# Metasurface-based planar microlenses for SPAD pixels

Jérôme Vaillant
*Univ. Grenoble Alpes*
*CEA, LETI*
38000 Grenoble, France
jerome.vaillant@cea.fr

Lucie Dilhan
*STMicroelectronics*
*Imaging division*
Grenoble, France
lucie.dilhan@st.com

Alain Ostrovsky
*STMicroelectronics*
*TR&D*
Crolles, France
alain.ostrovsky@st.com

Quentin Abadie
*Univ. Grenoble Alpes*
*CEA, LETI*
38000 Grenoble, France
quentin.abadie@cea.fr

Lilian Masarotto
*Univ. Grenoble Alpes*
*CEA, LETI*
38000 Grenoble, France
lilian.masarotto@cea.fr

Romain Paquet
*Univ. Grenoble Alpes*
*CEA, LETI*
38000 Grenoble, France
romain.paquet@cea.fr

Mickaël Cavelier
*Univ. Grenoble Alpes*
*CEA, LETI*
38000 Grenoble, France
mickael.cavelier@cea.fr

Cyril Bellegarde
*Univ. Grenoble Alpes*
*CEA, LETI*
38000 Grenoble, France
cyril.bellegarde@cea.fr

*Abstract*— In this paper we present two design generations of metasurface-based planar microlenses implemented on Front-Side Illumination SPAD pixels. This kind of microlens is an alternative to conventional reflow microlens. It offers more degrees of freedom in term of design, especially the capability to design off-axis microlens to gather light around the SPAD photodiode. The two generations of microlenses have been fabricated on STMicroelectronics SPAD and characterized. We validated the sensitivity improvement offered by extended metasurface-based microlens. We also confirmed the impact of lithography capability on metasurface performances, highlighting the need have access to advance deep-UV lithography.

## I. Introduction

Planar microlenses are well suited for SPAD pixels, which usually work under monochromatic illumination (Time-of-Flight or Fluorescence-Lifetime applications) and exhibit rather large pixel size with lower fill-factor compared to state-of-the-art CMOS pixels. We previously demonstrated Fresnel Zone Plate lenses[1] and proposed metasurface based[2] microlenses[3], [4]. In this paper, we present the design, fabrication and characterization results of two generations of metasurface-based planar microlenses on STMicroelectronics SPAD[5] array.

## II. META-ATOM AND PLANAR MICROLENS DESIGN

Our unitary structure, or meta-atom, is a nanoscale pillar of high refractive index material (amorphous silicon) embedded in a low refractive index medium (silicon oxide). The phase shift induced by a pillar is controlled by its geometry which is defined by the pitch, the capping thickness and the pillar's parameters: height and diameter (see Fig. 1).

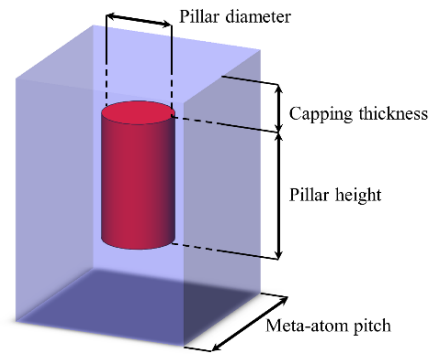

Fig. 1. Geometry of meta-atom defined by the meta-atom pitch, the thickness of silicon oxide capping and the pillar height and diameter

We have also considered the paving strategy. For the first generation of metasurface-based microlens, we considered square paving for meta-atom arrangement. In order to improve the spatial sampling, triangular paving (see Fig. 2) was also implemented in the second generation.

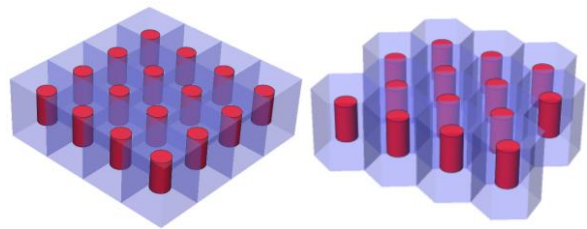

Fig. 2. Paving geometries used to design metasurface: square paving (left) and triangular paving (right)

We defined pillar libraries as sets of pillar with the paving and the first three parameters fixed and the pillar's diameter varies inside the range achievable by lithography: the minimal diameter is defined by the minimal Critical Dimension (CD) and the maximal diameter is defined as meta-atom pitch minus the minimal space. Among all possible libraries, we select the ones that cover a phase shift from 0 to $2\pi$ while offering the best transmission (see Fig. 3).

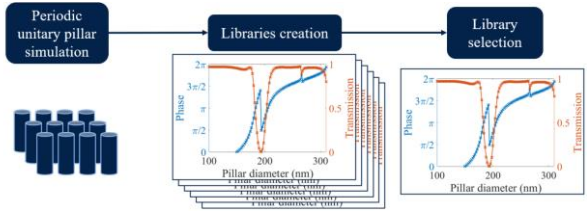

Fig. 3. Library generation and selection workflow

First generation of libraries is based on 500nm and 420nm pitches whereas the second generation is more aggressive with meta-atom pitch of 370nm. In both case, the minimum CD and space considered are 100nm.

### III. PLANAR MICROLENS DESIGNS

#### A. SPAD pixel

For this development, we consider $32 \times 32$ SPAD arrays. SPAD are sharing N-well and thus are grouped $4 \times 4$ into a $86.4 \times 86.4 \mu m^2$ cell (see Fig. 4). The SPAD itself has a dimension of $10.5 \times 11.5 \mu m^2$. So, we use the capability to design off-axis microlens with metasurface to extend the footprint compared to conventional refractive reflow-based microlens (i.e. $10.5 \times 11.5 \mu m^2$) and thus collect more light.

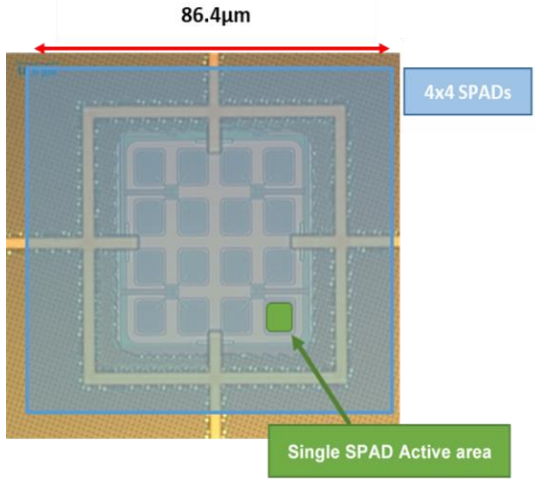

Fig. 4. Layout of 4x4 SPAD cell sharing the same N-well

#### B. Planar microlens

We divide the $32 \times 32$ array of SPAD into 8 areas of $8 \times 16$ SPAD. Each area is covered by a given design of microlens (see Fig. 5):

- One area without any microlens, to get the bare SPAD sensitivity as reference
- One area with a microlens having the same footprint as the reflow one ( $S_1 = 10.5 \times 11.5 \mu m^2$ ). This will be used for direct comparison with refractive microlens.
- Three area with microlens having intermediate footprint $S_{2.25} = 16.5 \times 16.5 \mu m^2$, which is 2.25 times larger than the reference surface.
- Three area with microlens having the largest possible footprint $S_{3.86} = 21.6 \times 21.6 \mu m^2$, 3.86 times larger than the reference.

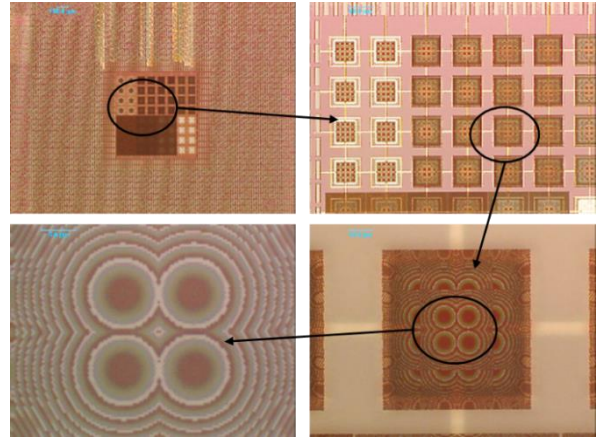

Fig. 5. Views of 32x32 SPAD array coverd by microlenses, with a focus on microlens footprint x2.25

It should be noted that for microlenses with footprint larger than the reference, the design differs according to their position inside the 4x4 group of SPAD (see bottom left picture in Fig. 5). Considering layout symmetry, we distinguish four different pixels: center, top, corner and side (see Fig. 6).

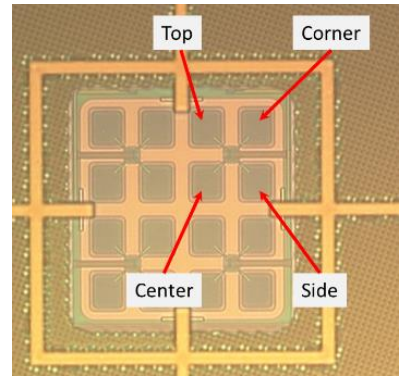

Fig. 6. Naming of SPAD and microlens according to their position in the 4x4 group

Microlens optical axis is centered on the SPAD of interest. When microlens and SPAD have the same footprint the optical axis is centered on it. But when the microlens surface extends outside SPAD footprint, the optical axis is offset (see bottom left image on Fig. 5). The target phase profile of the metasurface corresponds to a perfect lens bending a plane wave into a spherical one:

$$\varphi(x,y) = \frac{2\pi}{\lambda}\sqrt{(x-x_0)^2 + (y-y_0)^2 + f^2} - f \quad (1)$$

Spatial coordinates are denoted *x* and *y*, the offset of the optical axis of the microlens $x_0$ and $y_0$, the focal length of the microlens *f*, and the wavelength of interest λ (940nm in our case). To encode this phase profile we consider the classical look-up table method[4].

## IV. METASURFACE FABRICATION

The process flow starts on *40nm* CMOS Front-Side Illumination SPAD wafers[5] with the optical pedestal ($SiO_2$) deposition and planarization. Then, a low-stress layer of amorphous silicon (aSi) is deposited and planarized. The meta-atom are defined by dry deep-UV lithography and etching. Finally, $SiO_2$ deposition and planarization ensure the pillar's encapsulation and the capping thickness is tuned to minimize the reflection of the metasurface (see Fig. 7).

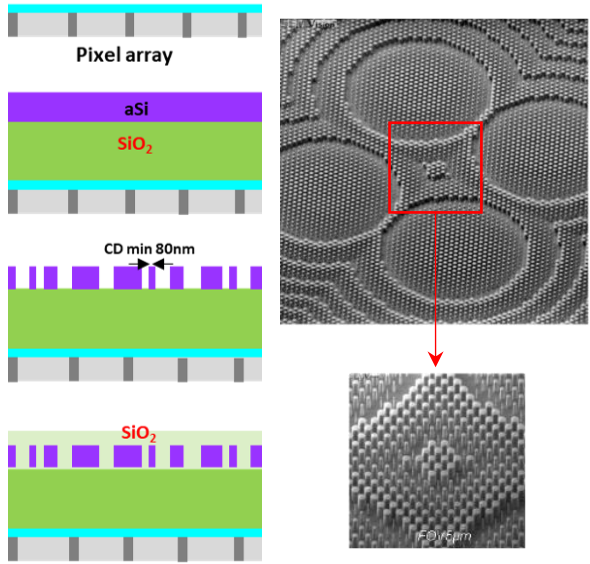

Fig. 7. Schematic of process flow (left) and SEM tilted view of planar metasurface based-microlens (right)

## V. ELECTRO-OPTICAL CHARACTERIZATION

### A. Experimental setup

The electro-optical characterization of the SPAD arrays are done at wafer-level with dedicated probe station based on Accretech 300mm prober. The light source is a Thorlabs M940L3 LED filtered with a Thorlabs bandpass filter FBH940-10 ($FWHM = 10nm$). Uniform illumination over the array is ensured by a LabSphere integrating sphere. Fixed distance between the sphere output and the wafer emulate an f/10 angular distribution. The light intensity is recorded using calibrated radiometer (UDT 221).

To evaluate microlens performances, we calculate the Photon Detection Efficiency (PDE), i.e. the quantum efficiency:

$$PDE = \frac{LCR-DCR}{\Phi_{940nm} \times a_{SPAD}^2} \quad (2)$$

With $\Phi_{940nm}$ the optical flux in $photon.m^{-2}.s^{-1}$ at the surface of the wafer, $a_{SPAD}^2$ the surface of the SPAD pixel (i.e. $21.6 \times 21.6 \mu m^2$), LCR the Light Count Rate and DCR the Dark Count Rate, which are the frequency of SPAD triggering respectively under illumination and in darkness. As these count rates usually depend of the excess bias, above the breakdown voltage of the SPAD, we first evaluate the mean breakdown voltage of each circuit. The excess bias is set to 1.5V above the breakdown voltage for LCR and DCR measurement.

### B. Experimental results

Both generation of metasurface-based microlens have been characterized as well as SPAD with refractive microlens (process of reference) and bare SPAD without microlens. The Fig. 9 shows the PDE for all of these configurations.

As expected, for bare SPAD and for microlens with unitary surface ($S_1 = 10.5 \times 11.5 \mu m^2$) the PDE are almost the same whatever the SPAD. Indeed, the layout of the 4 kind of SPAD (center, top, corner, side) are almost the same over this surface $S_1$. In this configuration, the refractive microlens leads to higher sensitivity than metasurface-based microlens as it bends the phase in a continuous way, contrary to metasurfaces that sample the phase profile, spatially and in phase value.

For extended surfaces ($S_{2.25}$ and $S_{3.86}$), we clearly see that the PDE depends on the position (center, top, corner, side): the central ones (dots in Fig. 9) are more sensitive than the corner ones (diamons in Fig. 9). There are two root causes to this effect. Firstly, the layout under a given microlens depends on its position: corner microlens is mainly above metallic interconnections that surround the $4 \times 4$ photodiodes, and layout below top and side microlenses are not exactly identical. Secondly, for the phase given by (1), the greater the distance to optical axis, the greater the slope of the phase. So as the meta-atom spatially sample the phase, this leads to possible aliasing. With the more aggressive design rules (smaller pitch and CD and use of triangular paving) the second generation of metasurfaces have higher and less dispersed performances.

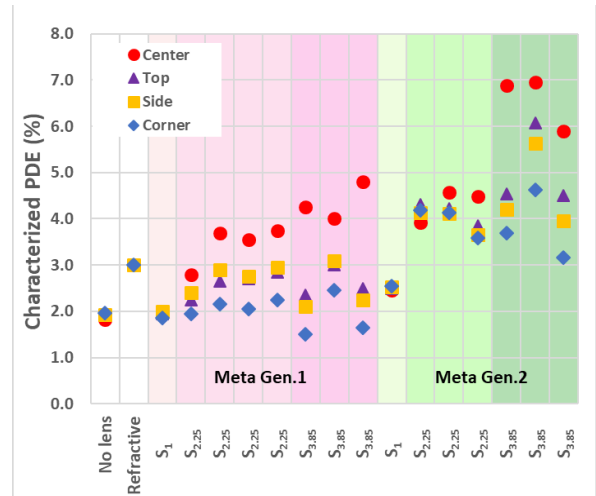

Fig. 8. PDE measurement for for bare SPAD (No lens), with refractive microlens (Refractive) and metasurface-based microlens with footprint of $S_1$, $S_{2.25}$ and $S_{3.86}$

Compared to reflow microlens, the PDE is improved whatever the metasurface-based microlens. The sensitivity is improved by 30% with small dispersion for our first extended design. For the largest microlens ($S_{3.86}$) we improve the PDE by × 2.3 for the center microlens, and in any case the PDE is higher than reference (reflow microlens).

## VI. Conclusion

This work validates the interest and the feasibility of metasurface-based microlens at pixel level. We demonstrate our capability to process deep-subwavelength pillars of amorphous silicon encapsulated in silicon dioxide on top of Front-Side Illumination CMOS wafer to generate microlenses. Measurement on 32x32 SPAD array confirms the interest of such technology. With the capability to design off-axis microlenses, we take advantage of space available around the 4x4 SPAD group to improve the PDE compared to the classical reflow microlenses.

## Acknowledgment

The authors would like to thanks STMicroelectronics and CEA-Leti people involved in the fabrication of CMOS wafers and metasurface-based microlens.